**2005 International Linear Collider Workshop – Stanford, U.S.A.**

# Electroweak Corrections for the Study of the Higgs Potential at the LC

F. Boudjema
*LAPTH, BP.110, Annecy-le-Vieux F-74941, France*

J.Fujimoto, T.Ishikawa, T.Kaneko, Y.Kurihara, Y.Shimizu
*High Energy Accelerator Research Organization(KEK), Tsukuba, Ibaraki 305-0801, Japan*

K.Kato
*Kogakuin University, Nishi-Shinjuku 1-24, Shinjuku, Tokyo 163-8677, Japan*

Y.Yasui
*Tokyo Management College, 625-1 Futamata, Ichikawa, Chiba 272-0001, Japan*

The full electroweak radiative correction is calculated for the process $e^+e^- \to \nu\bar{\nu}HH$ which is a window for the study of the Higgs potential at the future linear collider. The calculation is done by using GRACE, the automated system for the calculation of Feynman diagrams. The magnitude of the weak correction in the $G_\mu$ scheme is small in the high energy region where this process dominates over the *ZHH* production.
## 1. INTRODUCTION

The $e^+e^-$ linear collider (LC) will be a great asset for high-energy physics. One of the most important targets of the LC is the study of the Higgs particle. In Fig.1, we show various channels that are required for a study of the properties of the Higgs. Most of these channels have 3 or 4 particles in the final state. For instance, the process $e^+e^- \to ZH$ is only important at low energies, like at LEP2. It is quickly overcome by the process $e^+e^- \to \nu\bar{\nu}H$ at higher energies typical of the LC[1]. Probing the structure of the Higgs potential is achieved through the 3- and 4-body final states $e^+e^- \to ZHH$ and $e^+e^- \to \nu\bar{\nu}HH$ which involve the Higgs self coupling. For the analyses of experimental data with high accuracy, we need the radiative corrections(RC) for these processes in the standard electroweak(EW) theory. Though the RC for the multi-body final states is technically very complicated, the importance of these processes warrants a theoretical effort. A key solution for this task is the use of automated systems. The breakthrough was the computation of the complete EW RC for $e^+e^- \to \nu\bar{\nu}H$ in September 2002[2]. Since then other channels have been tackled by several groups[3-10]. In this talk, the results for the EW RC for $e^+e^- \to \nu\bar{\nu}HH$ is presented[11]. This completes the RC for Higgs production in the major channels for the future LC.

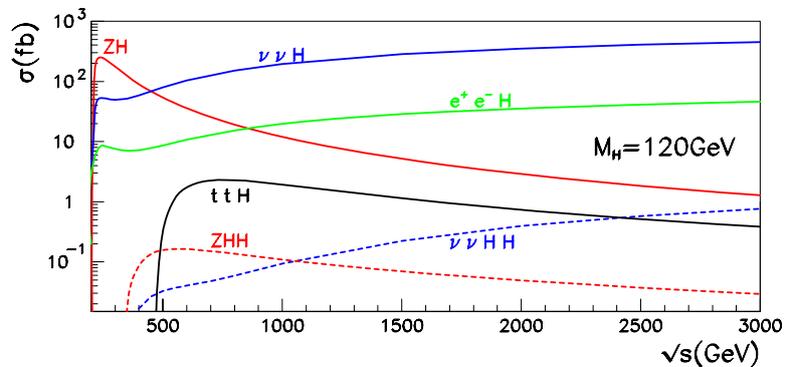

Figure 1: Tree-level cross sections for Higgs production at the LC.

## 2. RESULTS

We have already studied the RC for *ZHH* production in [7]. However, the extension to $\nu\bar{\nu}HH$ production is mandatory, since the latter cross section is larger than the former in the high energy region as shown in Fig.1. The $\nu\bar{\nu}HH$ production is more complicated and we need the reduction of 6-point functions. In the *ZHH* process, the *s*-channel mechanism, i.e., $e^+e^-$ annihilates into Z-boson, dominates while in the $\nu\bar{\nu}HH$ process the *WW*-fusion, which we call *t*-channel mechanism, is dominant at the high energy.

The calculation is done through the extensive use of GRACE, the automated system for the evaluation of Feynman diagrams[12]. The system generates the corresponding Feynman diagrams for the specified process. For $e^+e^- \to \nu_e\bar{\nu}_e HH$, the number of tree (one-loop) diagrams is 81(19638) and that for $e^+e^- \to \nu_\mu\bar{\nu}_\mu HH$ is 27(8292). (When we denote $e^+e^- \to \nu\bar{\nu}HH$, it stands for the sum over the three generations of neutrinos.) The diagrams are produced based on the EW theory with the non-linear gauge(NLG) fixing[12, 13]. The introduction of the NLG provides a systematic check over the numerical calculation. The number of diagrams quoted above refers to the full set of diagrams that are used for the gauge invariance check. At the the *production* or evaluation stage, a smaller set of diagrams is used, obtained after discarding the coupling between the scalars and the electron. The number of diagrams for the production set is 12(3416) for electron neutrino and 6(1754) for the muon/tau neutrino. Each product of a tree diagram and a one-loop diagram is processed by the symbolic manipulation program REDUCE or FORM to generate a FORTRAN code for that amplitude. The numerator including loop momenta is processed by a reduction method (see, e.g., [10]), so that the 5- and 6-point functions are converted into the sum of 4-point functions (and lower-point functions). These loop integrals are then fed into the FF loop integrals library[14]. Some of the box integrals including infrared divergence are calculated by an in-house library.

The parameter set for the calculation is as follows: $M_W$=80.4163GeV, $M_Z$=91.1876GeV, $\Gamma_Z$=2.4952GeV, $M_H$=120 GeV, $m_t$=180GeV, $W$=2 $E_{beam}$=400 ~ 2000GeV, $k_{cut}$=0.05$E_{beam}$. The width of the Z only appears at the resonant poles. The corresponding value of $\Delta r$ =0.022674.

We define the weak correction factor $\delta_W$ in the conventional way by subtraction of the QED correction from the full EW correction (see, e.g., [7]). The correction in the $G\mu$-scheme, $\delta_W^G$, is defined as $\delta_W^G = \delta_W - 4\Delta r$.

We have computed $\delta_W^G$ for $e^+e^- \to \nu_\mu\bar{\nu}_\mu HH$ and found that the values are similar to those for $e^+e^- \to ZHH$ shown in Fig.3 of [7]. [1] In Fig.2 and Fig.3, we show the calculated values of $\delta_W$ and $\delta_W^G$ for $e^+e^- \to \nu_e\bar{\nu}_e HH$ and for $e^+e^- \to \nu\bar{\nu}HH$. The preliminary results for the former was reported in [11]. [2] The energy dependence differs from that in *ZHH* channel, because the *t*-channel mechanism is dominant in the high energy region. The correction $\delta_W^G$ around 1 TeV and at higher energy is less than 5%. The experimental observation of such a correction will be difficult since the magnitude of cross section is $O(1fb)$ or less.

---

[1] The exact comparison is not possible, as the mass of top-quark is 174GeV in [7].
[2] The plotted values are updated from those presented in the talk at LCWS05. We realized that the output files of GRACE were not properly combined (*manually!*) for the figures of the LCWS05 presentation.

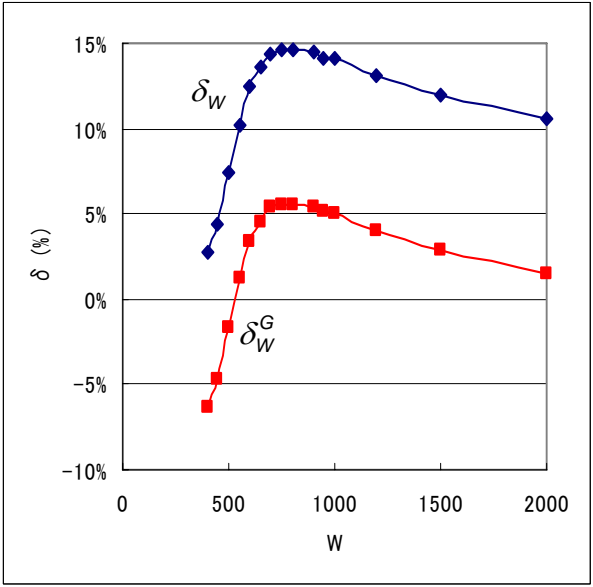

Figure 2: The weak correction for
$e^+e^- \to \nu_e \bar{\nu}_e HH$

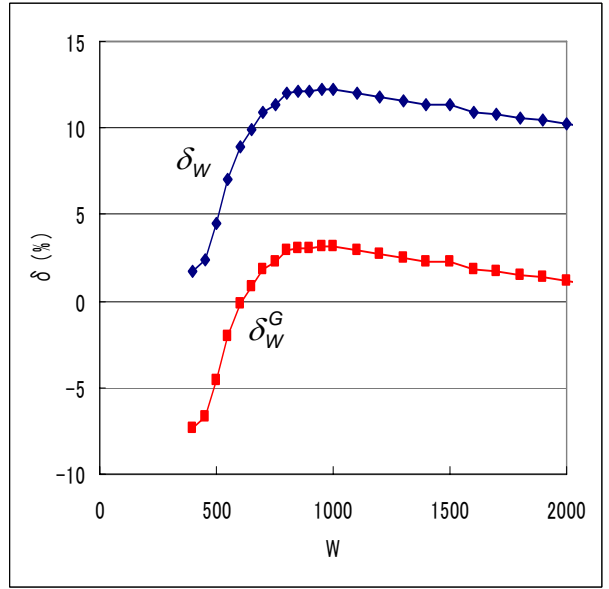

Figure 3: The weak correction for
$e^+e^- \to \nu \bar{\nu} HH$

## 3. WW→HH

The process $e^+e^- \to \nu\bar{\nu}HH$ consists of *s*-channel and *t*-channel diagrams. In the high energy region, the *t*-channel mechanism dominates, so that the study of the process $WW \to HH$ could provide the base for an approximation to $e^+e^- \to \nu\bar{\nu}HH$. At tree-level, 6 diagrams describe $WW \to HH$. They are classified into A-, B-, and C-terms, i.e., **A**nnihilation, **B**oson-exchange, and **C**ontact terms. The effect of the Higgs potential, i.e., the self-coupling of the Higgs, is included in the A-term. It should be noted that the sign of the amplitude of B-term is opposite to that of A-and C-terms.

The method of computation and the input parameters are the same as in $e^+e^- \to \nu\bar{\nu}HH$. The result is shown in Fig.4. In order to check the consistency between this result and that for $e^+e^- \to \nu\bar{\nu}HH$, we have considered the following approximation for $\delta_W$ for $e^+e^- \to \nu\bar{\nu}HH$.

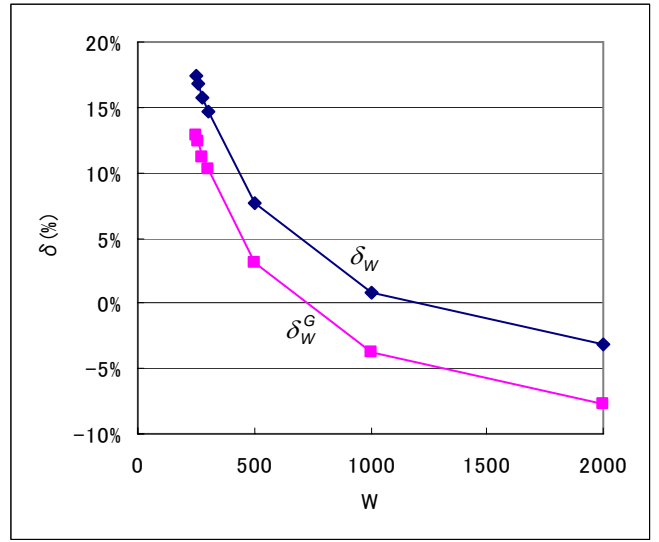

Figure 4: The weak correction to WW→HH.

$$\delta_W \approx \int F_\nu(M)\sigma_{WW}^{tree}(M)\delta_W^{WW}(M)dM / \int F_\nu(M)\sigma_{WW}^{tree}(M)dM + 2\Delta r \quad (1)$$

where $F_\nu(M)$, $\sigma_{WW}^{tree}$ and $\delta_W^{WW}$ are the distribution of invariant mass of Higgs pairs in $e^+e^- \to \nu\bar{\nu}HH$, the tree cross section of $WW \to HH$, and the $\delta_W$ for $WW \to HH$. In the energy region above 1TeV, this approximation reproduces $\delta_W$ for $e^+e^- \to \nu\bar{\nu}HH$ within 1%.

Sometimes the leading $m_t$ formula is used as an approximation to the RC to the Higgs vertices. The leading $m_t$ corrections for the triple Higgs vertex, WWH vertex and WWHH vertex are as follows:

$$C_H = -\frac{\alpha N_C}{3\pi s_W^2}\frac{m_t^4}{M_W^2 M_H^2} \quad C_W = -\frac{\alpha N_C}{8\pi s_W^2}\frac{m_t^2}{M_W^2}\frac{2s_W^2+3}{12s_W^2} \quad C_4 = -\frac{\alpha N_C}{8\pi s_W^2}\frac{m_t^2}{M_W^2}\frac{8s_W^2+3}{6s_W^2} \quad (2)$$

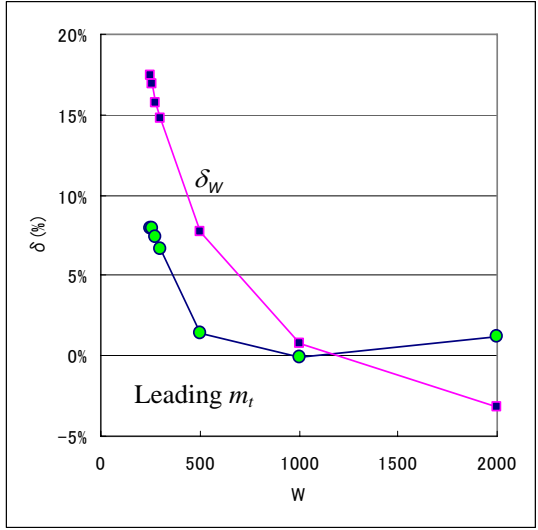

Figure 5: The exact weak corrections compared to those obtained with the leading $m_t$ approximation for $WW \to HH$.

Here, the correction factor is defined so that the corrected vertex is $(1+C)$ times the vertex at tree-level. The corrections for A-, B-, and C-terms are $C_H+C_W$, $2C_W$, and $C_4$, respectively. The numerical values of the C's in Eq.(2) are -0.11796, -0.02535, and -0.07034, respectively. If we take an artificially large value for $m_t$, say 100 times the physical value, the loop correction computed by GRACE is well described by this formula. However, with the physical value, i.e., $m_t$ =180GeV, the value of the loop amplitude differs from that given by these $m_t$ formulae. This is because in the real world the masses of Higgs and gauge bosons are not so small compared with $m_t$, and the energy is quite larger than $m_t$. In Fig.5, we show the comparison between the exact $\delta_W$ for $WW \to HH$ and that by the factors given above Eq. (2).

## 4. FINAL REMARKS

We have studied the RC for $e^+e^- \to \nu\bar{\nu}HH$ in the full EW theory. The energy dependence of the correction differs from that for $e^+e^- \to ZHH$. The magnitude of the weak correction $\delta_W^G$ with $M_H$=120GeV is only a few percent in most of the energy region where this process is important. Although this is a large scale computation, the GRACE system has performed well and given information and results only an exact calculation can provide. The study of RC for $WW \to HH$ is also done here and it is shown that this is the central mechanism in the TeV region. Also, the validity and the limitation of the leading $m_t$ approximation is presented through a comparison with the full computation.

With the results in this paper, the EW RCs for all processes shown in Fig.1 have been calculated. These results are indispensable for Higgs studies at the future LC.

PSN0601

## Acknowledgments

This work was supported in part by the Japan Society for Promotion of Science under Grant-in-Aid for Scientific Research B(No.14340081). This research is carried out under the auspices of Automated Calculations in Particle Physics (ACPP) - an International Research Group unit comprised of researchers from France, Russia and Japan (GDRI 397 of the French CNRS). We also thank IDRIS, *Institut du Dévelopement et des Ressources en Informatique Scientifique*, for the use of their computing resources.